\documentclass[english]{article}
\usepackage[T1]{fontenc}
\usepackage[latin9]{inputenc}

\usepackage{graphics}
\usepackage{graphicx}
\usepackage{epsfig}

\usepackage[textwidth=120mm,textheight=200mm]{geometry}

\usepackage{units}
\makeatletter

\providecommand{\tabularnewline}{\\}

\makeatother

\usepackage{babel}

\begin{document}

\title{Systematics of radial excitations in heavy-light hadrons}

\author{R. M. Woloshyn\\TRIUMF, 4004 Wesbrook Mall, Vancouver BC
V6T 2A3 Canada \and Mark Wurtz\\ Department of Physics and Astronomy\\ 
York University, Toronto, ON M3J 1P3 Canada }

\maketitle
\begin{abstract}
Some simple expectations for the quark mass dependence of radial excitation
energies of heavy-light hadrons based on consideration of nonrelativistic
quantum mechanics are discussed. Experimental and theoretical results
are reviewed in light of these expectations. Some new lattice QCD results
for masses of $\Lambda_b$ and $\Sigma_b$ baryons are presented.

\newpage{}
\end{abstract}

\section{Introduction}

The calculation of excited state masses in lattice QCD is quite challenging.
In recent years, using large operator bases and advanced analysis
methods some results for radial excitation energies of hadrons containing
a single heavy (charm or bottom) quark have been obtained. Assessing
these calculations is also not easy since they are often done without
the continuum limit or physical quark mass extrapolation having been
made. As well, experimental information about radial excitations of
heavy-light hadrons is very fragmentary. In any case, simply making
a number-by-number comparison of different results may not provide
the most insight into the physics of these systems. Some heuristics
that enable one to see global qualitative trends may be more informative
than individual number comparisons.

In this note we discuss some simple {}``rules of thumb'' for the
behaviour of radial excitation energies in different heavy-light systems.
These are motivated by consideration of the nonrelativistic quark
model and are obtained making severe simplifying assumptions. However,
the rules of thumb need not be exact. Rather, they serve
to focus our attention on the questions we should be asking as we
compare different calculations with each other and with experimental
data.

In Sect. 2 expectation for the quark mass dependence of radial excitation
energies of heavy-light hadrons is discussed using nonrelativistic
quantum mechanics as a guide. Experimental information on heavy-light
excitation energies is reviewed in Sect. 3. A sample of quark model
calculations are discussed in Sect. 4. These serve to assess the validity
of the rules of thumb presented in Sect. 2 and to provide a comparison
to the lattice QCD results reviewed in Sect. 5. A lattice QCD calculation 
for $\Lambda_b$ and $\Sigma_b$ baryons employing a free-form smearing method 
is outlined in the Appendix.

\section{Scaling in nonrelativistic quantum mechanics}

In this section we review the behaviour of excitation energy as a function
of constituent mass in the framework of nonrelativsitic quantum mechanics.
This is then used to develop some {}``rules of thumb'' for how heavy-light
hadron masses may be expected to depend on quark masses. 

The scaling argument follows the quark model discussions of 
Ref. \cite{Quigg:1977dd,Feldman:1978si,Quigg:1979v}.
Consider a simple power law potential $V(r)=cr^{\nu},$ the Schr\"odinger
equation for S-waves can be written in the form\begin{equation}
-\frac{1}{2\mu}\frac{d^{2}u}{dr^{2}}+cr^{\nu}u=Eu\end{equation}
 where $\mu$ is the reduced mass. Multiply by $2\mu$ and rescale
$r$ using \begin{equation}
\rho=\frac{r}{(2\mu c)^{p}}.\end{equation}
 The goal is to find $p$ such that all the explicit $\mu$ dependence can be
moved to the right hand side of (1). Using (2) it is easy to find
the necessary condition \begin{equation}
1+2p+p\nu=0\end{equation}
 which gives $p=\nicefrac{-1}{(\nu+2)}.$ Under these manipulations the
energy $E$ goes to $2\mu(2\mu c)^{2p}E.$ Since this combination of
factors must be independent of $\mu$ it implies that \begin{equation}
E\sim\frac{1}{\mu^{1+2p}}=\frac{1}{\mu^{\frac{\nu}{(\nu+2)}}}.\end{equation}
 In nonrelativistic quantum mechanics energy levels for a given system
can be shifted by a constant amount (by adding a constant to the potential)
but energy differences should obey (4) in any case. 

For a power law potential with $-2<\nu<0$ the excitation energy will
increase when $\mu$ increases while for a confining potential $\nu>0$
the excitation energy decreases with increasing $\mu.$ The common
quark model potentials have a Coulomb part ($\nu=-1)$ for which energy
would increase as $\mu$ and a linear confining part $(\nu=1)$ which
gives energies proportional to $\mu^{-\nicefrac{1}{3}}.$

It will be assumed that in a heavy-light hadron that the hadron size
is sufficiently large that the confining part of the potential determines
the mass dependence of the excitation energy. Furthermore, to include
baryons in the discussion it is assumed that the two light quarks
within a singly-heavy baryon act as an effective diquark with constituent
mass greater than the constituent mass of a single light quark. Then
considering the reduced mass\begin{equation}
\mu=\frac{m_{q}}{1+\frac{m_{q}}{M_{Q}}},\end{equation}
 where $M_{Q}$ and $m_{q}$ are the heavy and light masses respectively,
we can get the following rules of thumb:
\begin{enumerate}
\item Keeping the light mass(es) fixed and increasing the heavy mass will
decrease the energy of the radial excitation. For example, the radial
excitation energy of a B meson will be smaller than that of a D meson.
\item Keeping heavy mass fixed and increasing the light quark mass, that
is, going from $u,d$ quarks to strange will decrease the excitation
energy.
\item The radial excitation energy of a singly-heavy baryon will be less
than that of a heavy-light meson containing the same heavy quark flavour.
For example, the excitation energy of $\Lambda_{c}$ will be less
than that of D. 
\end{enumerate}
If empirical data proves to be consistent with these rules of thumb
they would provide a useful guide with which to assess in a broad
way the results of calculations of the spectrum. A large discrepancy
with these rules would be a signal that a simple understanding of
the physics is not correct. How the rules are violated may provide
some clues of where to search for the correct explanation.

\section{Experimental results}

At present there are no observed candidates for radial excitations of 
heavy-light hadrons 
with a b quark so rule no. 1 can not be tested empirically.

In the charm meson sector, the BaBar Collaboration has observed candidates
for the radial excitations of the $D^{0}$, $D^{*0}$, and $D^{*+}$
mesons\cite{delAmoSanchez:2010vq}. 
Using these results the radial excitation energies are calculated
including isospin averages where data are available and are given in
Table I. As well as individual pseudoscalar and vector meson excitation
energies the value using the spin averaged mass $\overline{M}=(M(J=0)+3M(J=1))/4$
is given for the $D$ meson. In the charm-strange sector an excited state with
$J^{\pi}=1^{-}$ has been observed\cite{Aaij:2012pc,pdg} 
which we interpret as the radial
excitation of $D_{s}^{*}$. As yet there is no identified candidate
for the radial excitation of $D_{s}.$ Comparing the values of $\Delta E(D^{*0})$
and $\Delta E(D_{s}^{*})$ one sees that the rule of thumb no. 2
is at best very weakly satisfied. 

\begin{table}

\caption{Experimental values for radial excitation energies.}

\centering{}\begin{tabular}{cc}
\hline 
 & Excitation energy{[}MeV{]}\tabularnewline
\hline
\hline 
$\Delta E(D^{0})$ & 674.6(8.2)\tabularnewline
$\Delta E(D^{*0})$ & 601.7(3.0)\tabularnewline
$\Delta E(D^{*\pm})$ & 611.0(4.7)\tabularnewline
$\Delta E(D^{*})$ & 606.4(5.6)\tabularnewline
$\overline{\Delta E(D^{0})}$ & 619.9(4.9)\tabularnewline
 & \tabularnewline
$\Delta E(D_{s}^{*})$ & 597(4)\tabularnewline
 & \tabularnewline
$\Delta E(\Lambda_{c}(2765))$ & 480(4)\tabularnewline
$\Delta E(\Lambda_{c}(2940))$ & 652.9(1.5)\tabularnewline
$\Delta E(\Xi_{c}(2980))$ & 502(4)\tabularnewline
\hline
\end{tabular}%
\end{table}

In the charm baryon sector no excited states have been positively identified
as radial excitations. The PDG\cite{pdg} particle listings show five excited
states of $\Lambda_{c}.$ Three of these have $J^{\pi}\neq\nicefrac{1}{2}^{+}.$
The excitation energies of the other two states, for which spin and
parity are undetermined, are given in Table 1. 
Chen \emph{et al.}\cite{Chen:2014nyo} advocate
the identification of $\Lambda_{c}(2765)$ with the radial excitation.
The PDG also list five $\Xi_{c}$ states
whose spins and parities are unknown\cite{pdg}. 
Chen \emph{et~al.} suggest that $\Xi_{c}(2980)$ is 
the first radial excitation\cite{Chen:2014nyo}. The excitation energy
for this state is noted in Table 1. The identification of radial 
excitations from \cite{Chen:2014nyo} would be consistent 
with the expectation that baryon excitation
energies are smaller than those of mesons (rule of thumb no. 3) and 
with quark model calculations as will be seen in the next section.
It would also suggest that rule of thumb 2 is weakly violated
by charmed baryons. 

\section{Quark models}

In this section some results from quark models are presented. The purpose
here is not to review the myriad of such calculations that have been
done but to show a sample of results that will serve as a point
of comparison and contrast to the lattice QCD results to be discussed
in the next section.

Table 2 shows calculated excitation energies for charmed mesons. The
results from different calculations (done over a period of about 25
years) are fairly consistent with Ref. \cite{DiPierro:2001uu} perhaps 
showing a significant
variation in the vector meson channel. The results are in reasonable
agreement with experimental values. There is no appreciable difference
in these models between the $D$ and the $D_{s}$ systems. Rule no.
2 is certainly not satisfied so likely the assumptions made were overly
simple. All these calculations incorporate some relativistic effects.
As well, the assumption that the confining potential dominates in
determining the light quark mass dependence may not be adequate here.

\begin{table}

\caption{Radial excitation energies in MeV of charm mesons from some quark model
calculations. The overline indicates a weighted average of pseudoscalar
and vector meson values.}

\centering{}\begin{tabular}{ccccc}
\hline 
 & Ref. \cite{Godfrey:1985xj} & Ref. \cite{Lahde:1999ih} & Ref. \cite{DiPierro:2001uu} & Ref.\cite{Ebert:2009ua}\tabularnewline
\hline
\hline 
$\Delta E(D)$ & 700 & 666 & 721 & 710\tabularnewline
$\Delta E(D^{*})$ & 600 & 595 & 685 & 622\tabularnewline
$\overline{\Delta E(D)}$ & 625 & 613 & 694 & 644\tabularnewline
 &  &  &  & \tabularnewline
$\Delta E(D_{s})$ & 690 & 689 & 731 & 720\tabularnewline
$\Delta E(D_{s}^{*})$ & 600 & 595 & 694 & 620\tabularnewline
$\overline{\Delta E(D_{s})}$ & 622 & 613 & 703 & 645\tabularnewline
\hline
\end{tabular}%
\end{table}

\begin{table}

\caption{Radial excitation energies in MeV of bottom mesons from some quark model
calculations. The overline indicates a weighted average of pseudoscalar
and vector meson values.}

\centering{}\begin{tabular}{ccccc}
\hline 
 & Ref. \cite{Godfrey:1985xj} & Ref. \cite{Lahde:1999ih} & Ref. \cite{DiPierro:2001uu} & Ref.\cite{Ebert:2009ua}\tabularnewline
\hline
\hline 
$\Delta E(B)$ & 590 & 545 & 607 & 610\tabularnewline
$\Delta E(B^{*})$ & 560 & 523 & 596 & 580\tabularnewline
$\overline{\Delta E(B)}$ & 567 & 529 & 599 & 588\tabularnewline
 &  &  &  & \tabularnewline
$\Delta E(B_{s})$ & 590 & 573 & 612 & 604\tabularnewline
$\Delta E(B_{s}^{*})$ & 560 & 549 & 598 & 573\tabularnewline
$\overline{\Delta E(B_{s})}$ & 567 & 555 & 602 & 596\tabularnewline
\hline
\end{tabular}%
\end{table}
 
Table 3 gives bottom meson results. The different models are reasonably
consistent and, as in the charm sector, rule no. 2 for the light quark
mass dependence is not evident. Comparing the results of Table 2 and
Table 3 one sees very clearly the heavy quark mass dependence expected
from rule no. 1.

The radial excitation energies for singly-heavy baryons calculated
in some potential models are listed in Table 4. As expected, the excitation
energy for bottom baryons is less than for charm baryons. Also, baryonic
excitation energies are smaller than mesonic ones. The calculated
values of $\Delta E(\Lambda_{c})$ support the identification of $\Lambda_{c}(2765)$
as a radial excitation as mentioned in Sect. 4.

\begin{table}
\caption{Radial excitation energies in MeV of singly-heavy baryons from some quark
model calculations.}

\centering{}\begin{tabular}{ccccccc}
\hline 
 & Ref. \cite{Capstick:1986bm}& Ref. \cite{Migura:2006ep}& Ref. \cite{Garcilazo:2007eh} & 
Ref. \cite{Roberts:2007ni}& Ref. \cite{Ebert:2011kk}& Ref. \cite{Yoshida:2015tia}\tabularnewline
\hline
\hline 
$\Delta E(\Lambda_{c})$ & 510 & 497 & 377 & 523 & 483  & 572\tabularnewline
$\Delta E(\Sigma_{c})$ & 450 & 488 & 345 & 503 & 458 & 569\tabularnewline
$\Delta E(\Xi_{c})$ &  &  &  & & 483 & \tabularnewline
 &  &  & & & & \tabularnewline
$\Delta E(\Lambda_{b})$ & 460 &  & 372 & 495 & 466 & 535\tabularnewline
$\Delta E(\Sigma_{b})$ & 405 &  & 338 &  461 & 405 & 520\tabularnewline
$\Delta E(\Xi_{b})$ &  &  &  &  & 463 &\tabularnewline
\hline
\end{tabular}%
\end{table}

\section{Lattice QCD simulations}

The calculation of excited state energies within the framework of
lattice QCD is very challenging. However, in the past decade large
scale simulations using large operator bases, variational methods
and advanced analysis techniques have started to yield results. There
is also difficulty in comparing different lattice simulations with
each other, with calculations done in other approaches and with experiment.
Lattice simulations are carried out with a variety of different discretized
actions. Often they are done on single gauge field ensembles \emph{i.e.,
}without continuum extrapolation and at light ($u,d$) quark masses
that are considerably larger that physical. Notwithstanding these
limitations, we believe it is worthwhile to assess the state of lattice
QCD simulations for heavy-light hadron radial excitations in view
of the rules of thumb proposed in Sect. 2 and of the results reviewed
in Sect. 3 and 4.

\begin{table}

\caption{Radial excitation energies in MeV of heavy-light mesons from recent
lattice QCD calculations.}

\centering{}\begin{tabular}{cccccc}
\hline 
 & Ref. \cite{Mohler:2012na} & Ref. \cite{Mohler:2011ke} & Ref. \cite{Kawanai:2015tga} & Ref. \cite{Wurtz:2015mqa} & Ref. \cite{Bernardoni:2015nqa}\tabularnewline
\hline
\hline 
$\Delta E(D)$ & 697(26) &  &  &  & \tabularnewline
$\Delta E(D^{*})$ & 642(29) &  &  &  & \tabularnewline
$\overline{\Delta E(D)}$ & 655(26) &  &  &  & \tabularnewline
 &  &  &  &  & \tabularnewline
$\Delta E(D_{s})$ &  & 754(27)(10) & 766(39)(50) &  & \tabularnewline
$\Delta E(D_{s}^{*})$ &  & 693(21)(10) & 719(43)(80) &  & \tabularnewline
$\overline{\Delta E(D_{s})}$ &  & 708(20)(10) & 731(41)(73) &  & \tabularnewline
\hline
$\Delta E(B)$ &  &  &  & 617(42) & 791(93)\tabularnewline
$\Delta E(B^{*})$ &  &  &  & 636(39) & \tabularnewline
$\overline{\Delta E(B)}$ &  &  &  & 632(38) & \tabularnewline
 &  &  &  &  & \tabularnewline
$\Delta E(B_{s})$ &  &  &  & 594(14) & 566(57)\tabularnewline
$\Delta E(B_{s}^{*})$ &  &  &  & 595(15) & \tabularnewline
$\overline{\Delta E(B_{s})}$ &  &  &  & 595(15) & \tabularnewline
\hline
\end{tabular}%
\end{table}

Table 5 summarizes results of recent lattice simulations for heavy-light
meson excitation energies. The calculations in Ref. \cite{Mohler:2012na} 
and Ref. \cite{Mohler:2011ke} were
done using different gauge field ensembles and different quark masses
so these results do not inform us about the $D$ versus $D_{s}$ comparison
suggested by rule no. 2. Ref. \cite{Mohler:2011ke} and Ref. \cite{Kawanai:2015tga} used the same gauge
field ensemble but the calculational methods were completely different.
However, the results are completely compatible. In Ref. \cite{Kawanai:2015tga} 
a potential
was first determined and this was then used in the Schr\"odinger
equation to calculate energies. Ref. \cite{Mohler:2011ke} used the standard 
technique
of extracting masses from the Euclidean time dependence of meson two-point
functions. Comparison of these charm meson results with Table 1 shows
that lattice QCD values to be somewhat larger than experimental values.
Comparing to Table 2, one sees that lattice QCD results are not inconsistent
with potential model calculations.

The calculations in Ref. \cite{Mohler:2011ke} and Ref. \cite{Wurtz:2015mqa} were 
done using the same gauge field
ensemble although the lattices actions used for the heavy quark were
different, Fermilab clover and NRQCD for the charm and bottom quarks
respectively. The expected decrease
of the excitation energy in going from charm to bottom is clearly
exhibited. We note also that the lattice QCD results of Ref. \cite{Wurtz:2015mqa}
are fairly compatible with potential model calculations. 

The results of Ref. \cite{Bernardoni:2015nqa} are quite interesting. 
The value for $\Delta E(B)$
is determined rather poorly but it does appear to be somewhat of an
outlier. It doesn't quite fit into the pattern established by experiment,
potential models or other lattice QCD simulations. The difference
between $\Delta E(B)$ and $\Delta E(B_{s})$is large (although only
about $2\sigma$ significant) and the value of $\Delta E(B)$ seems
contrary to rule of thumb no. 1 which is satisfied in all other calculations.

\begin{table}
\caption{Radial excitation energies in MeV of heavy-light baryons from recent lattice
QCD calculations.}

\centering{}\begin{tabular}{cccc}
\hline 
 & Ref. \cite{Bali:2015lka} & Ref. \cite{Padmanath:2014bxa} & This work\tabularnewline
\hline
\hline 
$\Delta E(\Lambda_{c})$ & 903(76)(80) & 781(18) & \tabularnewline
$\Delta E(\Sigma_{c})$ & 676(95)(98) & 758(22) &\tabularnewline
$\Delta E(\Xi_{c})$ & 792(39)(46) & & \tabularnewline
&  &  &\tabularnewline
$\Delta E(\Lambda_{b})$ &  &  & 740(92)\tabularnewline
$\Delta E(\Sigma_{b})$ &  &  & 643(60)\tabularnewline
\hline
\end{tabular}%
\end{table}

Results of recent lattice QCD calculations of excitation energies
of charm baryons are given in Table 6. The calculations of 
Ref. \cite{Bali:2015lka} were done
at a single lattice spacing, about 0.075fm, and extrapolated to physical 
quark masses. The calculations of Ref. \cite{Padmanath:2014bxa} were 
done with $u,d$ quarks corresponding to a
pion mass of about 379MeV on an anisotropic lattice with temporal and spatial
lattice spacings of 0.034fm and 0.12fm. Taken at face value they are at variance
with the expectation that baryon excitations are smaller than meson
excitations and with potential model calculations. They suggest quite
strongly that the $\Lambda_{c}(2765)$ is not the radial excitation
of $\Lambda_{c}$. Also shown are the results from an exploritory 
study of single bottom baryons using the same lattice setup as in
Ref. \cite{Wurtz:2015mqa} (see the Appendix). As in the charm
sector, the lattice simulation yields a bottom baryon excitation energy
larger than expected from quark models and larger than calculated for 
heavy-light mesons. 
Should these patterns persist with improvements
in lattice simulations and confirmation by new experimental information
that would present a significant challenge to our understanding
of heavy-light hadrons.

\section{Summary}

Some simple {}``rules of thumb'' for the quark mass dependence of
radial excitation energies, motivated by nonrelativistic quantum mechanics,
were proposed. Experimental results, quark models and lattice QCD
calculations were reviewed in light of these expectations. In 
quark models and lattice QCD bottom 
mesons have smaller excitation energies than charm mesons as
expected. This is not yet confirmed experimentally.
As well, in quark models baryons have smaller excitation 
energies than mesons containing the same heavy flavour. However, the
expectation that increasing the light quark mass, that is, going from
$u,d$ to strange should decrease excitation energy is not evident.

For the
most part, quark models and lattice QCD simulations give a pattern
of quark mass dependence in heavy-light mesons which is compatible with 
available experimental results.
The rules of thumb allow one to spot results which are possible outliers.
For example, the results of Ref. \cite{Bernardoni:2015nqa}
for excitation energy of $B$ and
$B_{s}$ do not fit the pattern established by other calculations.

The results of recent lattice QCD calculations
of heavy-light baryon excitation energies, Table 6, 
seem to be at variance with the expectation that they should
be smaller than excitation energies of heavy-light mesons. A confirmation
of this pattern would challenge the quark model as a guide to heavy-light
baryon spectroscopy.

We hope that this note will provide motivation for more study of excited
heavy-light hadrons to fill the gaps in experimental information
and in lattice QCD simulations.

\section*{Appendix}

Free-form smearing \cite{georg13} allows for the construction of correlation functions
selectively enhancing the contribution of particular states. A variation
of this method was shown to be quite effective in the 
calculation of the spectrum of bottomonium and B mesons\cite{Wurtz:2015mqa}.
Here we explore the efficacy of free-form smearing 
in the simulation of single bottom baryons.

The essential idea of free-form smearing is to start at a single lattice
source point $y$ and smear the quark field over the whole spatial
lattice volume using the reweighting formula
\begin{equation}
\hat{\psi}_{y}(x)=\frac{\tilde{\psi}_y(x)}{\left<\left|\left|\tilde{\psi}_y(x)\right|\right|\right>}f(x-y)
\end{equation}
 where $f(x-y)$ is an arbitrary profile function. The field $\tilde{\psi}_{y}(x)$
is obtained by starting at the source point $y$ and extending the
field $\psi(y)$ by multiplying by gauge field links following minimal
paths to all spatial sites in the source time slice
\begin{equation}
\tilde{\psi}_{y}(x)={\displaystyle \sum_{minimalpaths}U(x\rightarrow y)\psi(y).}
\end{equation}
The smeared fields $\hat{\psi}_{y}(x)$ are used as the source
quark fields in the construction of the hadron two-point functions.

In our study of mesons \cite{Wurtz:2015mqa} the profile functions were chosen to have the
shape of Coulomb wavefunctions, that is, $f(x-y)$ is  $G = e^{-\frac{r}{a_{0}}}$
and $E = e^{-\frac{r}{a_{0}}}(r-b)$ for S-wave ground and excited states
respectively where $r$ is the shortest distance between $y$ and
$x$ in a periodic box. The range $a_{0}$ and node position $b$
can be adjusted to improve the isolation of ground and excited states.

In this work we explore the application of free-form smearing to the
calculation of spin-$\nicefrac{1}{2}$ $\Lambda_{b}$ and $\Sigma_{b}$
baryon masses. The bottom quark is described using lattice NRQCD as
in Ref. \cite{Wurtz:2015mqa} and is not smeared in the correlation function 
construction. The light quarks are simulated with the clover action 
using code from the DD-HMC package \cite{ddhmc} and can 
be unsmeared or smeared at the source using either a ground state($G$) or
an excited state($E$) profile function. This yields six correlation functions
with different source smearing for each baryon interpolating operator.
The baryon operators used in this work are
\begin{equation}
\Lambda_{b}=\frac{1}{\sqrt{6}}\epsilon^{abc}\{2[q_{a}^{T}C\gamma_{5}q'_{b}]Q_{c}+[q_{a}^{T}C\gamma_{5}Q_{b}]q'_{c}-[{q'}{}_{a}^{T}C\gamma_{5}Q_{b}]q_{c}\}
\end{equation}
and 
\begin{equation}
\Sigma_{b}=\epsilon^{abc}[q_{a}^{T}C\gamma_{5}q{}_{b}]Q_{c}
\end{equation}
where $Q$ is the heavy b-quark field and $q,q'$ are light ($u,d)$
fields. The so-called heavy or nonrelativistic lambda (see (3) 
in \cite{Mathur:2002ce}) 
was also considered but was not used in the final analysis. The
relativistic forms used here allow for both positive and negative parity
baryon states to be simulated.

It is natural to consider the heavy quark as acting approximately as a static
color source and to smear the light quarks about it.
Ideally one would like to explore baryon operators which incorporate
correlations between the light quarks to mimic, for example, a quark-diquark
structure as commonly used in quark model calculations of baryon spectra.
We do not attempt to do this here. Smearing was applied independently 
to each light quark. This may be a limitation of the present approach.

The lattice setup was the same as used in \cite{Wurtz:2015mqa}. An $N_f = 2 + 1$ flavour dynamical
gauge field ensemble from the PACS-CS Collaboration \cite{pacscs09}
was used. The lattice was
$32^3\times64$ with a lattice spacing of $0.0907(13)$fm determined by the 
PACS-CS Collaboration. The pion mass is $156(7)$ MeV for the light quarks used
in the simulation. Other parameters are described in \cite{Wurtz:2015mqa}.

Correlation functions for positive and negative parity baryons were calculated for 
198 gauge field configurations averaging over 16 source time positions for
each configuration. It was found that correlation functions without smearing
did not provide useful data. The effective simulation energies did not reach
a plateau before the signal disappeared into noise. Correlators with light 
quark source smearing profiles GG, EE, and GE were analyzed.

\begin{figure}

\centerline{
\includegraphics[width=150mm]{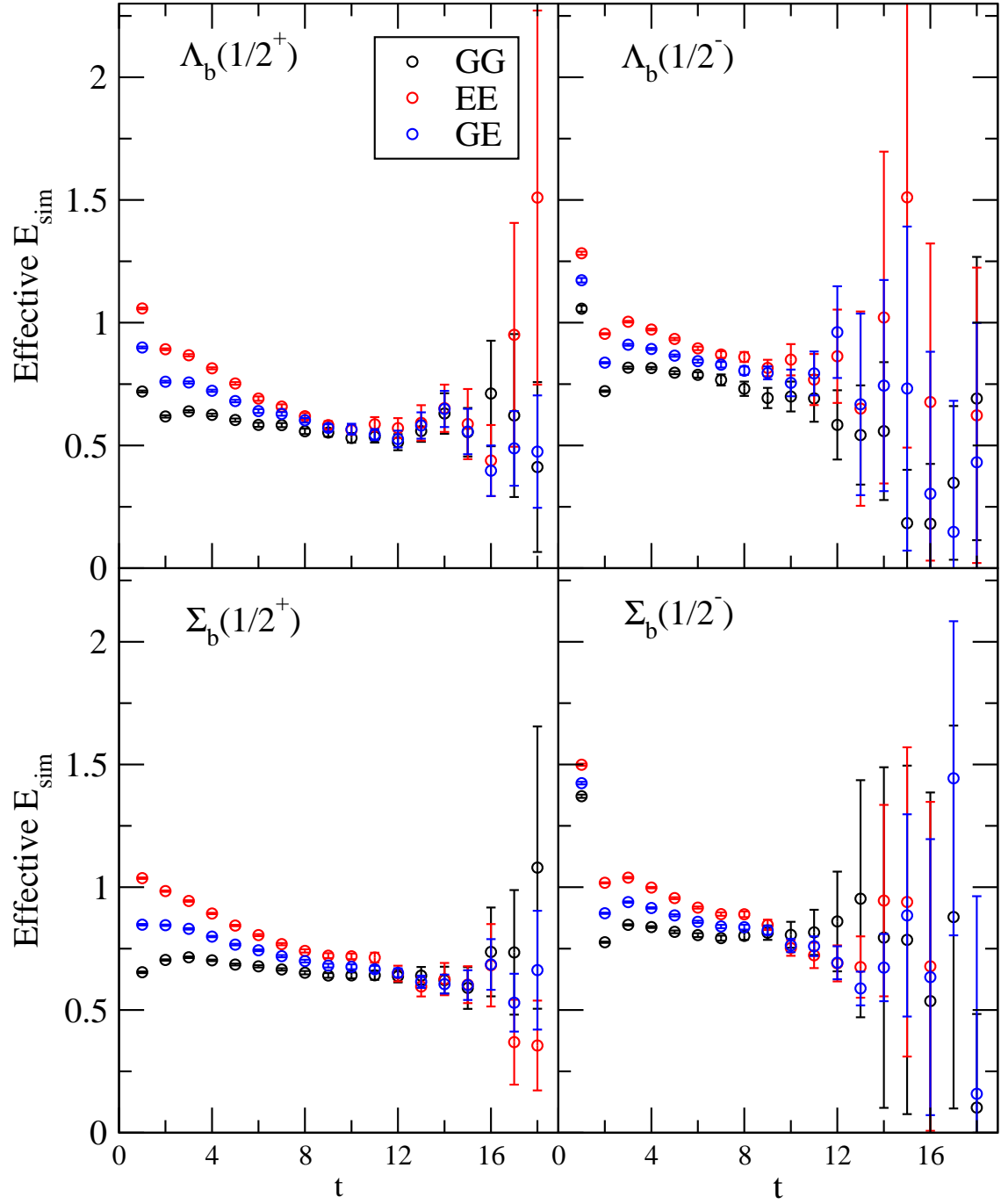}}
\caption{Effective simulation energies in lattice units.}
\label{fig_eff}

\end{figure}

\begin{figure}

\centerline{
\includegraphics[width=100mm]{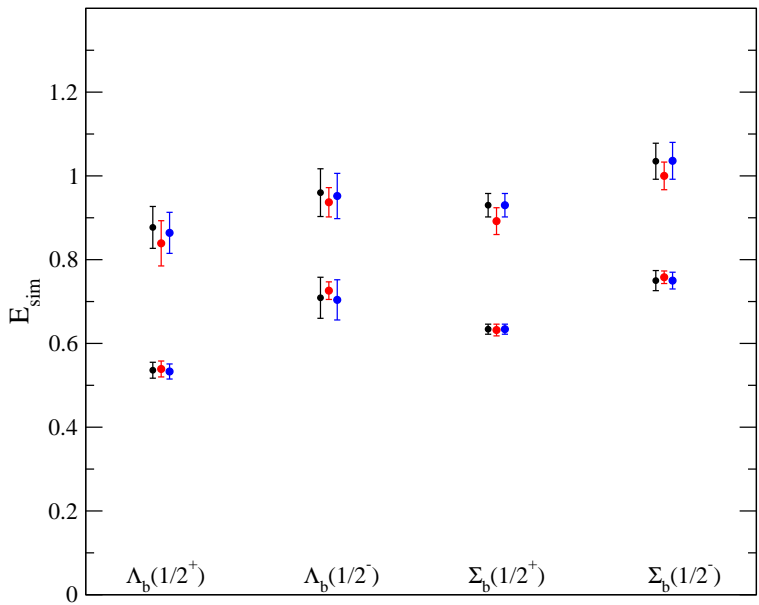}}
\caption{Simulation energies in lattice units for the two lowest states 
from different fits.}
\label{fig_esim}

\end{figure}

Figure \ref{fig_eff} shows the effective simulation energies. A variety of
constrained multi-exponential fits were done. For the positive parity
correlators points up to $t = 15$ were included. For negative parity
corelators only times up to 11 were used. The results were quite robust
with regard to initial time (2 or 3), number of exponential terms (3 or 4) 
and choice of priors. Representative fit values for ground and first 
excited state simulation energies are shown in Fig.~\ref{fig_esim}.

The mass difference ${\Sigma}_b(1/2^+) - {\Lambda}_b(1/2^+)$ was found to be 
213(42) MeV consistent with the experimental value\cite{pdg} 194(3) MeV. The
mass difference ${\Lambda}_b(1/2^-) - {\Lambda}_b(1/2^+)$ is poorly 
determined 344(105) MeV compared to the experimental value\cite{pdg} 293(1) MeV. 
For ${\Sigma}_b(1/2^-) - {\Sigma}_b(1/2^+)$ we have a prediction 
of 252(60)MeV.
The radial excitation energies of the positive parity states are 
given in Table 6.

\section*{Acknowledgment}

We thank R.~Lewis for a discussion which prompted writing this note and for
helpful comments and N.~Mathur for providing some results from Ref. \cite{Padmanath:2014bxa}.


\begin{thebibliography}{99}

\bibitem{Quigg:1977dd}
C.~Quigg and J.~L.~Rosner.
Phys. Lett. {\bf 71B}, 153 (1977).

\bibitem{Feldman:1978si}
G.~Feldman, T.~Fulton and A.~Devoto,
Nucl. Phys. {\bf B514}, 441 (1978).

\bibitem{Quigg:1979v}
C.~Quigg and J.~L.~Rosner.
Phys. Rept. {\bf 56}, 167 (1979). 

\bibitem{delAmoSanchez:2010vq}
P.~del~Amo~Sanchez {\emph {et al.}} (BaBar Collaboration),
Phys. Rev. D {\bf 82}, 111101 (2010).

\bibitem{Aaij:2012pc}
R.~Aaij {\emph {et al.}} (LHCb Collaboration),
JHEP 1210, 151 (2012).

\bibitem{pdg}
K.~A.~Olive {\emph {et al}}. (Particle Data Group), 
Chin. Phys. C {\bf 38}, 090001 (2014).

\bibitem{Chen:2014nyo}
B.~Chen, K.-W.~Wei and A.~Zhang, 
Eur. Phys. J. A {\bf 51}, 82 (2015).

\bibitem{Godfrey:1985xj} 
S.~Godfrey and N.~Isgur,
Phys.\ Rev.\ D {\bf 32}, 189 (1985).

	
\bibitem{Lahde:1999ih}
T.~A.~L\"ahde, C.~J.~Nyf\"alt and D.~O.~Riska,
Nucl. Phys. {\bf A674},  114 (2000).

\bibitem{DiPierro:2001uu}
M.~Di~Pierro and E.~Eichten,
Phys. Rev. D {\bf 64}, 114004 (2001).

\bibitem{Ebert:2009ua}
D.~Ebert, R.~N.~Faustov and V.~O.~Galkin 
Eur. Phys. J. C {\bf 66}, 197 (2010).

\bibitem{Capstick:1986bm}
S.~Capstick and N.~Isgur,
Phys. Rev. D {\bf 34}, 2809 (1986).

\bibitem{Migura:2006ep}
S.~Migura, D.~Merten, B.~Metsch and H.-R.~Petry,
Eur. Phys. J. A {\bf 28}, 41 (2006).

\bibitem{Garcilazo:2007eh}
H.~Garcilazo, T.~Vijande and A.~Valcarce,
J. Phys. G {\bf 34}, 961 (2007).

\bibitem{Roberts:2007ni}
W.~Roberts and M.~Pervin,
Int. J. Mod. Phys. A {\bf 23}, 2817 (2008).

\bibitem{Ebert:2011kk}
D.~Ebert, R.~N.~Faustov and V.~O.~Galkin, 
Phys.Rev. D {\bf 84}, 014025 (2011).

\bibitem{Yoshida:2015tia}
T.~Yoshida, E.~Hiyama, A.~Hosaka, M.~Oka and K.~Sadato,
arXiv:1510.01067[hep-ph].

\bibitem{Mohler:2012na}
D.~Mohler, S.~Prelovsek and R.~M.~Woloshyn,
Phys. Rev. D {\bf 87}, 034501 (2013).

\bibitem{Mohler:2011ke}
D.~Mohler and R.~M.~Woloshyn,
Phys. Rev. D {\bf 84}, 054505 (2011).

\bibitem{Kawanai:2015tga}
T.~Kawanai and S.~Sasaki,
Phys. Rev. D  {\bf 92}, 094503 (2015).
 
\bibitem{Wurtz:2015mqa}
M.~Wurtz, R.~Lewis and R.~M.~Woloshyn,
Phys. Rev. D {\bf 92}, 054504 (2015).

\bibitem{Bernardoni:2015nqa}
F.~Bernardoni, B.~Blossier, J.~Bulava, M.~Della Morte, 
P.~Fritzsch, N.~Garron, A.~G\'erardin and J.~Heitger {\emph{et al.}},
Phys. Rev. D {\bf 92}, 054509 (2015).

\bibitem{Bali:2015lka}
P.~Pérez-Rubio, S.~Collins and G.~S.~Bali,
Phys. Rev. D {\bf 92}, 034504 (2015).

\bibitem{Padmanath:2014bxa}
M.~Padmanath, R.~G.~Edwards, N.~Mathur and M~J.~Peardon,
PoS LATTICE2014 084 (2015).

\bibitem{georg13}
G.~M.~von Hippel, B.~J\"ager, T.~D.~Rae and H.~Wittig,
JHEP {\bf 1309}, 14 (2013).

\bibitem{ddhmc}
M.~L\"uscher,
Comput. Phys. Commun. {\bf 165}, 199 (2005).

\bibitem{Mathur:2002ce}
N.~Mathur, R.~Lewis and R.~M.~Woloshyn
Phys. Rev. D {\bf 66}, 014502 (2002).
  
\bibitem{pacscs09}
S.~Aoki, {\emph et al.} [PACS-CS Collaboration],
Phys. Rev. D {\bf 79}, 034503 (2009).
 
\end{thebibliography}
\end{document}